\def\lesssim{{_ <\atop{^\sim}}}

\def\ap3m{AP$^3$M}
\def\LCDM{$\Lambda$CDM}

\def\hkpc{$h^{-1}{\ }{\rm kpc}$}

\def\hMsun{$h^{-1}{\ }{\rm M_{\odot}}$}
\def\kms{${\rm{\ }km{\ }s^{-1}}$}
\def\nbody{$N$-body}
\def\c15{$c_{\rm 1/5}$}

\def\Rvir{$R_{\rm vir}$}

\newcommand{\Fig}[1]{Figure~\ref{#1}}
\newcommand{\mlapm}{\texttt{MLAPM}}

\def\ea{et~al.~}                            

\def\lesssim{\mathrel{\hbox{\rlap{\hbox{\lower4pt\hbox{$\sim$}}}\hbox{$<$}}}}
\def\gtrsim{\mathrel{\hbox{\rlap{\hbox{\lower4pt\hbox{$\sim$}}}\hbox{$>$}}}}

\newcommand{\AAA}[3]    {\mbox{#3, A\&A, #1, #2}}
\newcommand{\ApJ}[3]    {\mbox{#3, ApJ, #1, #2}}

\newcommand{\MNRAS}[3]  {\mbox{#3, MNRAS, #1, #2}}
\newcommand{\Nature}[3] {\mbox{#3, Nature, #1, #2}}
\newcommand{\astroph}[1]{\mbox{\texttt{astro-ph/#1}}}

\documentstyle[epsfig]{mn}

\begin{document}
\title[The outskirts of clusters]
{The evolution of substructure III: the outskirts of clusters}

\author[Gill~\ea]
       {Stuart P.~D. Gill$^{1}$,
			 Alexander Knebe$^{1,2}$, Brad K. Gibson$^1$ \\
        	$^1$Centre for Astrophysics~\& Supercomputing, 
            Swinburne University, Mail \#H39, P.O. Box 218, 
            Hawthorn, Victoria, 3122, Australia\\
        	$^2$Astrophysikalisches Institut Potsdam, 
         	An der Sternwarte 16, 14482 Potsdam, Germany\\
       	}
\date{Received ...; accepted ...}

\maketitle

\begin{abstract}
We present an investigation of satellite galaxies in the outskirts of
galaxy clusters taken from a series of high-resolution \nbody\
simulations. We focus on the so-called ``backsplash population'',
i.e. satellite galaxies that once were inside the virial radius of the
host but now reside beyond it. We find that this population is
significant in number and needs to be appreciated when interpreting
the various galaxy morphology environmental relationships and
decoupling the degeneracy between nature and nurture. Specifically, we
find that approximately half of the galaxies with current clustercentric distance in the interval 1 -- 2 virial radii of the host are
backsplash galaxies which once penetrated deep into the cluster
potential, with 90\% of these entering to within 50\% of the virial
radius. These galaxies have undergone significant tidal disruption,
loosing on average 40\% of their mass. This results in a mass function
for the backsplash population different to those galaxies infalling
for the first time. We further show that these two populations are kinematically
distinct and should be observable within existent spectroscopic
surveys.
\end{abstract}


\section{Introduction}

The relationship between galaxy morphology and local environment
(i.e. the morphology-density relation) was first noticed by Hubble~\&
Humason (1931), where they reported that field and cluster galaxy
populations differ. Oemler (1974) extended this finding by showing
that the relationship held for differing clusters richness. The field
truly emerged when Dressler (1980) demonstrated the strong
relationship over five orders of magnitude between the local density
of galaxies and the proportions of different morphological
types. Bhavsar (1981), de Souza et al. (1982), and Postman~\& Geller
(1984) extended this work further to include the environments of both
loose and compact groups. Recently, Aguerri~\ea (2004) performed a
thorough analysis of 116 bright galaxies in the Coma cluster, finding
that bluer objects are located at larger projected radii while
simultaneously showing a larger velocity dispersion than their red
counterparts. Moreover, the bluest objects also host the most
prominent disks contrary to systems observed close to the cluster
centre or in high-density environments.  Environmental dependence of
galactic stellar populations is also seen in the Butcher-Oemler effect
(Butcher \& Oemler 1978, Kodama \& Bower 2001) with clusters at higher
redshifts showing a greater fraction of blue objects than are seen at
present.

The above observational work supports the idea that galaxies in
clusters are substantially different from galaxies in the field.  But
the origins of these morphology-density relationships are still not
fully understood with several large and small scale mechanisms
proposed to explain their existence, including ram pressure stripping
(Gun \& Gott 1972), tidal stripping/star formation (Merritt 1983,
1984), starvation (Larson, Tinsley \& Caldwell 1980), galaxy merger
and harassment (Icke 1985; Moore \ea 1996) and dynamics (Tsuchiya \& Shimada 2000).

\begin{figure}
\centerline{\psfig{file=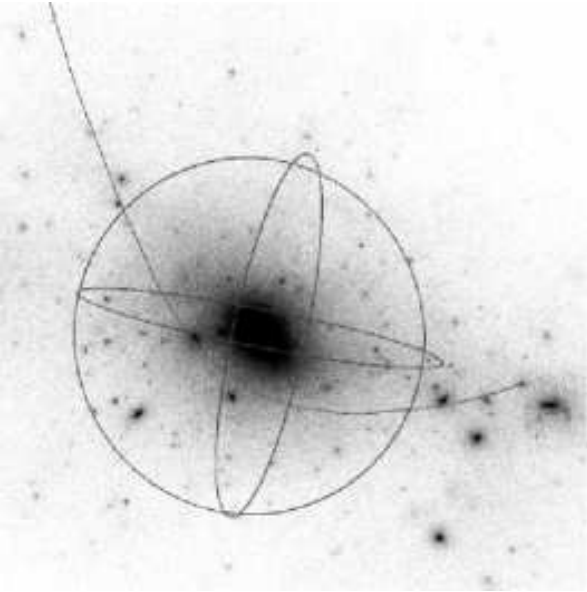,width=\hsize}}
 \caption{A simulated cluster at z=0 with the virial radius indicated by the dark sphere. The other line in this figure represents the orbital path of a ``backsplash'' galaxy. Such a galaxy has previously spent time near the cluster centre but now lies outside the virial radius of the cluster.}
 \label{galback}
\end{figure}

Recent cosmological simulations (both hydrodynamical and \nbody) have
provided valuable insight into the mechanisms responsible for the
morphology-density relationship (Springel \ea 2001; Goto \ea 2003; Okamoto \&
Nagashima 2003). In the analysis which follows, we focus on the
dynamics of satellite galaxies taken from a series of high-resolution,
fully self-consistent, cosmological simulations of eight galaxy
clusters.  We concentrate on the outskirts of these clusters, i.e.
distances in the range $[R_{vir},2.5 R_{vir}]$, which
(observationally) have only recently been probed through wide-field
optical imaging and spectroscopy (Miyazaki~\ea 2002; Lewis~\ea 2002).
We will demonstrate that a rich population of galaxies exist beyond
the virial radius most of which have previously spent time near the
cluster centre and can be seen in Figure~\ref{galback}. We characterise the spatial, velocity and mass
properties of this population and contrast these with those of the
spatially coincident newly infalling galaxies.

Our work complements the earlier studies of Balogh \ea (2000) and
Mamon \ea (2004). Balogh \ea investigated the \textit{particle} backsplash
from cosmological simulations and found that 50 $\pm$ 20\% of the
particles within $[R_{200},2R_{200}]$ had passed through the $R_{200}$
radius. Mamon \ea recently extended this work to calculate the maximum
backsplash distance for particles to be $2.5 R_{100}$ and for galaxies
$1.7 R_{100}$.

The outline of the paper is as follows. Section~\ref{Computation}
provides a description of the cosmological simulations employed.  We
investigate the number distribution of galaxies in the cluster
outskirts in Section~\ref{number}, with the mass distribution
discussed in Section~\ref{mass} and the velocity distribution of the
satellites investigated in Section~\ref{velocity}. We finish with our
summary and conclusions in Section~\ref{Conclusions}.

\section{Simulation Details}\label{Computation}

Our analysis is based on a suite of eight high-resolution \nbody\
simulations (Gill, Knebe~\& Gibson 2004a) carried out using the
publicly available adaptive mesh refinement code \mlapm\ (Knebe,
Green~\& Binney 2001) in a standard \LCDM\ cosmology ($\Omega_0 =
0.3,\Omega_\lambda = 0.7, \Omega_b h^2 = 0.04, h = 0.7, \sigma_8 =
0.9$). Each run focuses on the formation and evolution of a dark
matter galaxy cluster containing of order one million particles, with
mass resolution $1.6 \times 10^8$ \hMsun\ and force resolution 
$\sim$2\hkpc\ which is of the order 0.05\% of the host's virial
radius. These simulations have the required resolution to follow the
satellites within the very central regions of the host potential
($\geq$5--10\% of the virial radius) and the time resolution to resolve the satellite dynamics with good accuracy ($\Delta t \approx$170~Myrs).  Such temporal resolution provides of order 10-20 timesteps per orbit per satellite galaxy, Thus allowing these simulations to be used in a previous paper Gill \ea (2004b) to accurately measure the orbital parameters of each individual satellite galaxy. 

The clusters were chosen to sample a variety of environments and their
details are summarised in (Gill, Knebe~\& Gibson 2004a). We define the
virial radius $R_{\rm vir}$ as the point where the density of the host
(measured in terms of the cosmological background density $\rho_b$)
drops below the virial overdensity $\Delta_{\rm vir}=340$.
$\Delta_{\rm vir}=340$ is based upon the dissipationless spherical
top-hat collapse model and is a function of both cosmological model
and time.  We further applied a lower mass cut for all the satellite galaxies used
in this paper at $2 \times 10^{10}$ \hMsun\ (100 particles).  For further specific details of the host halos such as masses and density profiles please refer to the earlier papers in the series (cf. Gill, Knebe~\& Gibson 2004a and Gill \ea 2004b).

For later reference and comparison to previous studies by Balogh~\ea
(2000) and Mamon~\ea (2004) we relate $R_{vir}$ to other definitions
of the virial radius, namely $R_{200}$ and $R_{100}$. According to the
NFW profile (Navarro, Frenck~\& White 1997) the dark matter density in
the outer regions of halos drops like $\rho \propto r^{-3}$ and hence
$R_{vir} \sim 1.4 R_{200}$ and $R_{vir} \sim R_{100}$.

\section{Number distribution of galaxies in the cluster outskirts}\label{number}

The fact that galaxies in clusters are different from galaxies in the
field has been attributed to a variety of mechanisms acting on a
variety of scales. It has been suggested that galaxies can ``rebound''
up to $1.7 R_{100}$ (Mamon \ea 2004), thus blurring the definition of
``cluster'' and ``field''. In this section we investigate cluster galaxy
``backsplash'': how significant is the population of galaxies that once
passed through a cluster's virial radius but now resides in its
outskirts?

\begin{figure*}
\centerline{\psfig{file=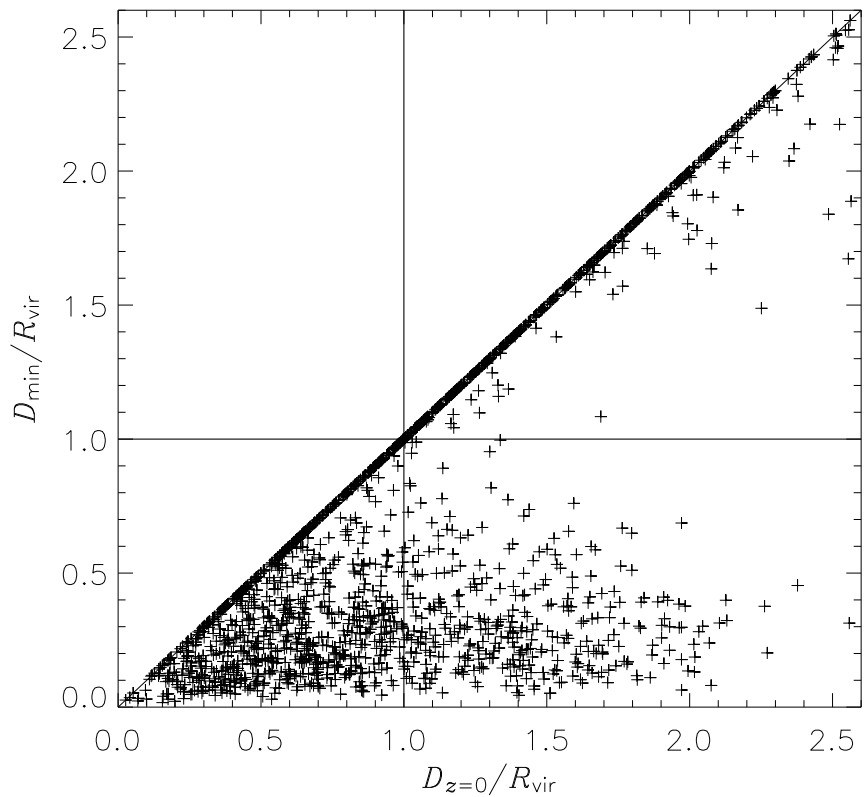,width=13cm}}
 \caption{Minimum distance for all satellite galaxies versus current distance in terms of the host's virial radius. The population in the upper right corner consists of substructure orbiting within satellites, i.e. sub-subhalos. This figure clearly indicates that there is a distinct population of ``backsplash'' satellites, i.e. $D_{\rm min} < R_{\rm vir}$ and $D_{z=0} > R_{\rm vir}$ in the lower right corner.}
 \label{RminRz}
\end{figure*}

In Figure~\ref{RminRz} we plot for all of our eight simulated clusters the minimum distance $D_{\rm min}$ a galaxy reached to the cluster centre throughout its history versus its current distance $D_{z=0}$. Both distances have been normalised by the cluster's present day virial radius $R_{\rm vir}$. There are four distinct populations of satellites visible in this figure:\\

\noindent
 \underline{1. $D_{\rm min} = D_{z=0}$:} \hfill (the infalling population)

\noindent
       These satellites are falling in for the first time.\\

\noindent
 \underline{2. $D_{\rm min} > R_{\rm vir}$ and $D_{z=0} > R_{\rm vir}$:} \hfill (infalling sub-population)

\noindent
       This population is made up of subhalos orbiting satellites,
       i.e. sub-subhalos.\\

\noindent
 \underline{3. $D_{\rm min} < R_{\rm vir}$ and $D_{z=0} > R_{\rm vir}$:} \hfill (backsplash satellites)

\noindent
       These satellites once passed through the virial radius
       of the respective host but have ``rebound'' to the outskirts
       of the galaxy cluster.\\

\noindent
 \underline{4. $D_{\rm min} < R_{\rm vir}$ and $D_{z=0} < R_{\rm vir}$:} \hfill (the bound population)

\noindent
       This is the ``normal'' satellite population orbiting within the virial radius and bound to the host.\\

\begin{figure}
\centerline{\psfig{file=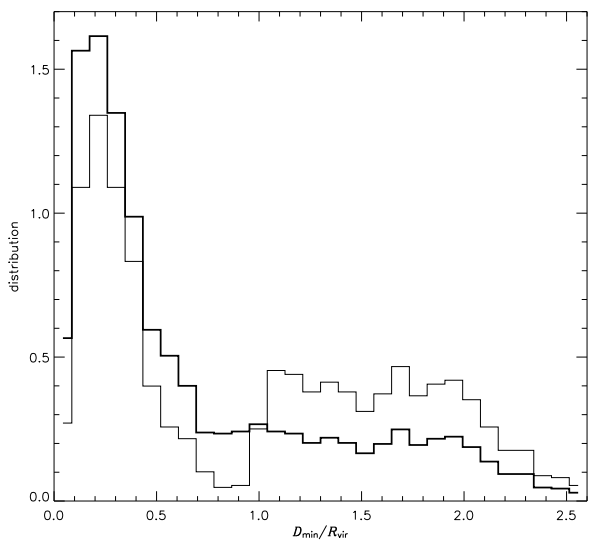,width=\hsize}}
 \caption{The relative normalised distribution function of the minimum distance $D_{\rm min}/R_{\rm vir}$ for the whole galaxy population (thick histograms) and the satellites with current positions greater than the virial radius (thin histograms). We note that they both peak near 25\% of the cluster virial radius.}
 \label{DminNdist}
\end{figure}

Our data sets indicates that the maximum backsplash distance encountered is $\sim 2.5 R_{vir}$ which is larger than the value presented in Mamon \ea (2004)\footnote{refer to Section~\ref{Computation} for conversion of $R_{\rm vir}$ to $R_{100}$.} yet not widely in disagreement, further very few backsplash galaxies exist beyond $\sim 2 R_{vir}$.

A more quantitative analysis reveals that 30\% of all satellites with
$D_{\rm min} < R_{\rm vir}$ are outside the host's virial radius at
$z=0$. Further, 50\% of all galaxies with a current distance $D_{z=0}$
in the range $[R_{\rm vir}, 2R_{\rm vir}]$ are in fact backsplash
galaxies, consistent with the value $50
\pm 20$ \% quoted by Balogh~\ea (2000) (based on cluster particles as
opposed to gravitationally bound satellites though).

Only 2\% of the backsplash population has had more than one orbit
(i.e. eight satellites in total) and most stem from single passages
through the host's virial radius, respectively. And each of these
eight ``multiple passage'' backsplash galaxies are found in the range
[$R_{\rm vir}$, 1.2$R_{\rm vir}$] which is close to the
host.

We now investigate the depth to which backsplash galaxies penetrate the cluster potential. To this extent we plot in Figure~\ref{DminNdist} the normalised number distribution of the minimum distance $D_{\rm min}$ for the entire satellite population (thick histogram) and the restricted set of satellites currently outside the virial radius (thin histograms). It is interesting to note that the distributions are fairly similar for $D_{\rm min} < R_{\rm vir}$, i.e. they both peak near 0.25. We stress that this is the minimum distance as measured over the lifetime of the host halo and hence is not to be confused with the last pericentre whose distribution peaks at about 35\% of $R_{\rm vir}$ (Gill~\ea 2004b). Figure~\ref{DminNdist} reveals that the a number of the galaxies that are now in the outskirts of the cluster have had passages as close to the host as their bound counterparts which implies that the former are on highly radial orbits.  This is consistent with the findings of Solanes \ea (2001) who found evidence that gas-poor spirals in HI deficient clusters move in orbits more radial than those of the gas-rich objects.  In fact, 90\% of our backsplash satellites once passed within the inner 50\% of the virial radius.

In summary the number of backsplash galaxies is significant and should be accounted for when interpreting the galaxy morphology-density relationship. These galaxies penetrate deep within the cluster potential, as deep as their bound counterparts. Hence, they should be sampling the large and small-scale transformation mechanisms alluded to in Section 1. For example, if we used the prescriptions outlined in Treu~\ea (2003), these galaxies would also have undergone starvation, ram pressure stripping, tidally triggered star formation and significant tidal stripping, thus confusing any correlation of the galaxy's morphology with local environment.  

\begin{figure}
\centerline{\psfig{file=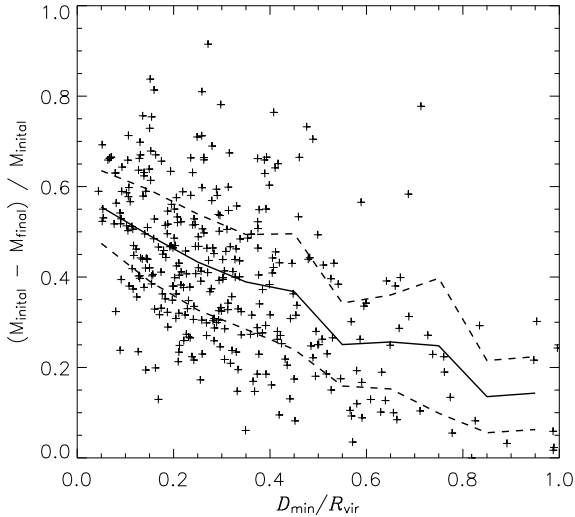,width=\hsize}}
 \caption{Relative mass loss for the backsplash satellites over the
          lifetime of the host as a function of minimum distance.
          The solid line shows the average with the dashed lines being
          the standard deviation.}
 \label{MassChRmin}
\end{figure}

\section{Mass distribution of the galaxies in the cluster outskirts}\label{mass}

One way to gauge the significance of these transformation mechanisms
is to investigate the mass loss of a galaxy during its encounter with
the cluster. To this extent we focus only on the backsplash population
and plot the relative mass lost over the lifetime of the host as a
function of minimum distance $D_{\rm min}/R_{\rm vir}$ for each of these
satellite in \Fig{MassChRmin}. We find that the closer a satellite
gets to the host the stronger the tidal stripping. This result is not
surprising but needs to be viewed from the perspective of the
backsplash population: galaxies in the outskirts experienced
significant tidal interactions and mass loss, in some cases up to 80\%
of their original mass.

Further to that we see little dependence of \textit{present day}
clustercentric distance on the mass lost (not presented though). At
all distances outside the virial radius the average mass lost for each
backsplash galaxy is $\sim$ 40\%.

\begin{figure}
\centerline{\psfig{file=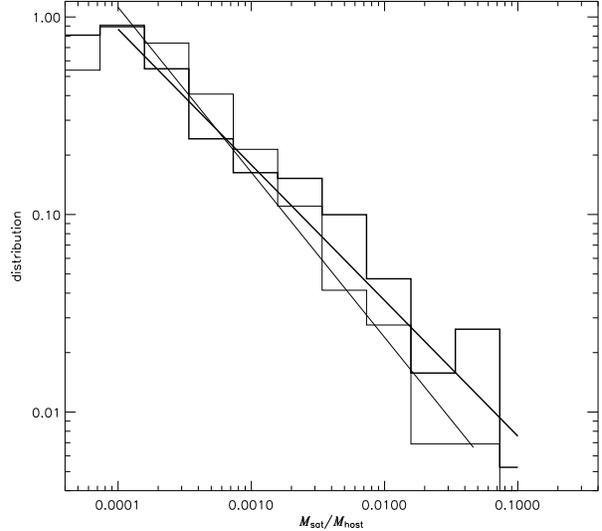,width=\hsize}}
 \caption{The relative number distribution of satellite masses. 
          The thin histogram represents the backsplash population
          and the thick histogram shows the satellites infalling for the first time.}
 \label{MassSpec}
\end{figure}

Since the population of backsplash galaxies has undergone significant
tidal stripping we expect the mass spectrum to be different from that
of the galaxies infalling for the first time. In Figure~\ref{MassSpec}
we plot the relative number distribution of satellites with a certain
mass $M_{\rm sat}$ measured in terms of the host's virial mass $M_{\rm
host}$. The thin histogram represents the backsplash population and
the thick histogram shows the infalling population.  To better
discriminate between these two populations we fitted the mass spectra
to a simple power-law

\begin{equation}
 n(M) = C M^{\alpha} \ .
\end{equation}

\noindent
There is a marginal difference between the slopes, the backsplash population
having $\alpha=0.9 \pm 0.3$ whereas the infalling satellites
distribution is characterised by $\alpha = 0.7 \pm 0.3$. The steeper
slope of the backsplashed galaxies reflects the lack of larger
satellites and an over-abundance of smaller galaxies,
respectively. This tilt in the mass spectrum is readily explained by
the mass loss highlighted in Figure~\ref{MassChRmin}. We note that the
first mass bin was not used in fitting the power-law as it simply
reflects the \textit{absolute} mass cut of $2 \times 10^{10}$
\hMsun\ (i.e. we did not use a relative mass cut with respect to the
mass of the host).

\section{Velocity properties of galaxies in the cluster outskirts}\label{velocity}

To populate the outskirts of clusters, galaxies that once passed
through the virial radius and close to the centre of the host must
have had high velocities. In this section we explore the relative
velocities of satellite galaxies within the cluster and its outskirts.

\subsection{A kinematically distinct backsplash population}\label{simudata}
\begin{figure*}
\centerline{\psfig{file=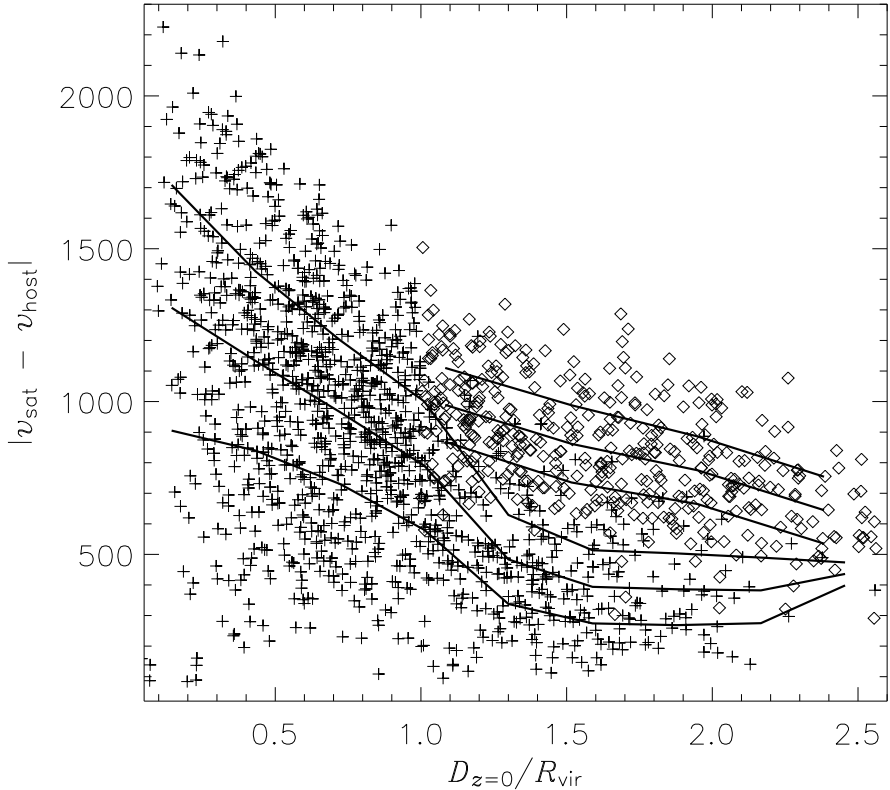,width=13cm}}
 \caption{The absolute value of the relative velocity for the infalling satellites
          (diamonds) and the population of satellite that have crossed the virial radius at sometime (crosses) as a function
          of current cluster distance to the host. The middle solid lines represent averages
          with the outer lines being the standard deviations.
          Note the kinematic distinction between the populations.}
 \label{Rvel}
\end{figure*}

Figure~\ref{Rvel} shows the absolute value of the relative velocity
between a satellite galaxy and the host cluster as a function of
current clustercentric distance. We present the results for two
populations again, namely the infalling satellites that are still
outside the virial radius (diamonds) and the backsplash galaxies
alongside galaxies within the virial radius (crosses).

At the virial radius the maximum relative velocity for a galaxy in our
data set is 1500\kms\ with no galaxy farther away having a greater
velocity. This finding can be used to explore the two competing models
for the ram-pressure stripping observed in NGC 4522, a spiral galaxy
in the Virgo cluster with a distance close to the virial radius
(Vollmer \ea 2001). Kenney~\ea (2004) suggested two possible scenarios
to explain its ongoing ICM-ISM stripping: either NGC 4522 is
experiencing stripping at this distance due to bulk motions and local
density enhancements of the ICM produced by shocks in the ICM, or --
since ram-pressure stripping is proportional to $v^2$ -- the galaxy
has a very high relative velocity, of order 4000\kms. As hinted by
Kenney et al., such a velocity is extremely unlikely, and as shown in
\Fig{Rvel} only in the very central regions of clusters do galaxies
reach these high relative velocities.
  
Figure~\ref{Rvel} furthermore shows that the backsplash population is
quite distinct kinematically; the infalling satellites have
significantly larger velocities than the backsplash galaxies.  This
provides a possible mechanism to observationally detect these rebound
satellites.

\begin{figure}
\centerline{\psfig{file=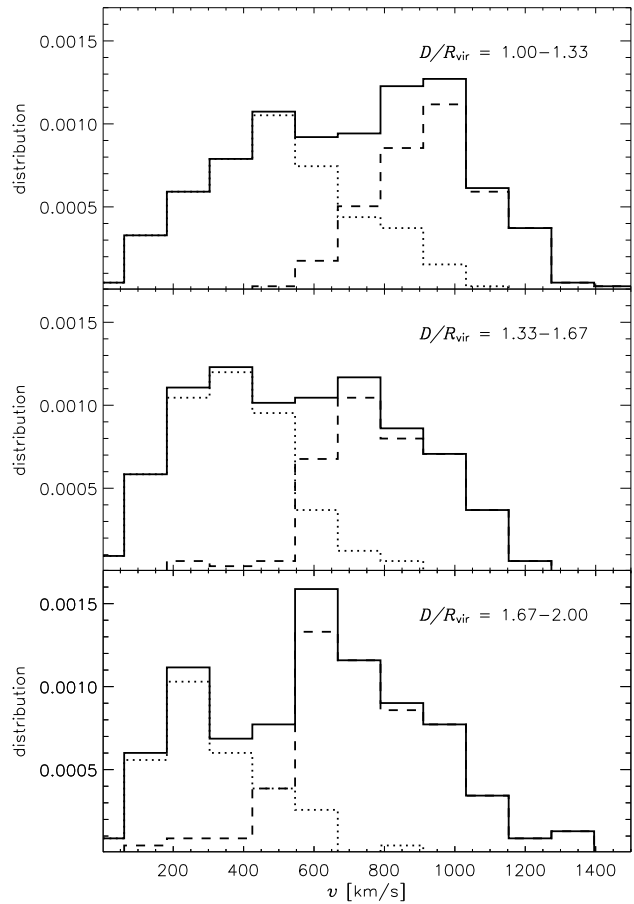,width=\hsize}}
 \caption{The distribution of the relative velocities
          in three distance bins, i.e. 1.00--1.33, 1.33--1.67 and 1.67--2.00 virial
          radii. The dotted histograms shows the backsplash galaxies while
          the dashed lines represent the infalling galaxies (both normalized to
          the total number of satellites). The solid histogram is the sum of both distributions. We see that the distribution has
          two distinct peaks.}
 \label{Nvrel}
\end{figure}

To investigate this in more detail we divided all (infalling and
backsplash) satellites into three radial bins between 1 and 2
\Rvir. In \Fig{Nvrel} we show the distribution of the relative
velocities for all satellites in the respective bin (solid lines).
The dotted (dashed) lines are for the backsplash (infalling)
population alone, normalised to the total number of satellites
(i.e. the actual distribution is simply the sum of these two
distributions). From the distributions presented in \Fig{Nvrel} it is
rather obvious that the backsplash population should be detectable by
simply plotting the velocity distribution function (VDF) for cluster
galaxies in the range $[R_{\rm vir},2R_{\rm vir}]$: there are two
distinct peaks with the lower velocity peak indicative of the rebound
satellites. Moreover, as we move further away from the host the
separation between the two populations becomes even more pronounced.

One might still pose the question though, if the backsplash population
can be identified by their ``flight path'', i.e. the orientation of
their velocity vector. In Table~\ref{Inouts} we therefore summarise the
numbers of backsplash galaxies moving towards (``approaching'') and
away (``receding'') from the centre of the host. This demonstrates
that the backsplash galaxies cannot simply be selected by the
direction of their velocity as they appear to move in both
directions. However, as distance increases we do find a higher
percentage of backsplash galaxies moving away from the host.

\begin{table}\label{Inouts}
\caption{Number of inbound and outbound backsplash galaxies in different
         radial bins.}
\begin{tabular}{ccc}\hline
distance & receding galaxies & approaching galaxies \\
\hline \hline
(1.00 - 1.33) \Rvir & 102 & 106 \\
(1.33 - 1.67) \Rvir & 87  & 57  \\
(1.67 - 2.00) \Rvir & 47  & 22  \\
\end{tabular}
\end{table}

\subsection{Observational impact} \label{Obs}

Thus far, we have relied upon the fact that our simulation data provides full six dimensional velocity and spatial information. In this section we extend the velocity distribution function to the ``observer's plane'' by restricting the data to projected distances and line-of-sight velocities; we make this transformation by placing the potential observer at infinity.

\begin{figure}
\centerline{\psfig{file=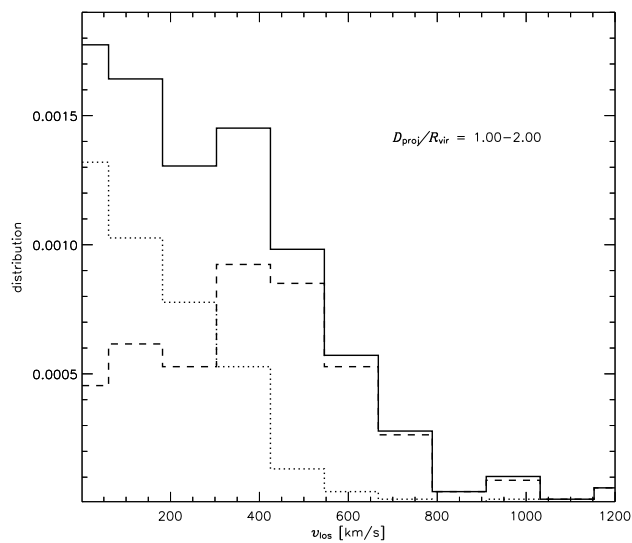,width=\hsize}}
 \caption{The distribution of the line-of-sight velocities (relative to the host) for galaxies with projected distance between 1.0--2.0 virial radii. Velocities have been convolved with the 2dF velocity uncertainty of 100\kms, typical of that encountered in multi-object spectroscopic surveys (e.g. Colless \ea 2001). The dotted histogram represents the backsplash galaxies while the dashed histogram is based upon the infalling galaxies (both normalized to the total number of satellites). The solid histogram is the sum of both distributions.}
 \label{Nvrel-obs}
\end{figure}

In Figure~\ref{Nvrel-obs} we show the distribution of the
line-of-sight velocities (still with respect to the host) for
galaxies with projected distances between 1.0 and 2.0 virial
radii. Velocities have also been convolved with a velocity
uncertainty of 100\kms, typical of that encountered in existing multi-object
spectroscopic surveys such as the 2dFGRS (Colless \ea 2001). The dotted histogram shows
the backsplash galaxies while the dashed histogram represents the
infalling galaxies (both normalised to the total number of
satellites). The solid histogram is the sum of both distributions.

Not surprisingly, this figure appears quite different from
Figure~\ref{Nvrel} due to the loss of information when transforming
from 6-dimensional space to 3-dimensional space. The kinematically
distinct bimodal populations are no longer readily separable.
However, if the backsplash population does not exist, one should
simply observe the dashed histogram in Figure~\ref{Nvrel-obs}. The
presence of a backsplash population ``distorts'' the Gaussian velocity
distribution function such that it increases with decreasing
line-of-sight velocity.

Figure~\ref{Nvrel-obs} should now be comparable to extant
cluster data sets such as the 2dFGRS (Lewis \ea 2002). Lewis \ea show
VDFs for 17 clusters (of order 50 galaxies per cluster). Stacking the
17 cluster VDFs, eliminating all galaxies within the (projected)
virial radius, should yield a composite 1--2 $R_{vir}$ VDF with $\sim
200-300$ galaxies; such a composite VDF should either support or
refute our predicted result (Figure~\ref{Nvrel-obs}). However, this is just a first step to a fair comparison. The next step would require colour, surface brightness and kinematic selection.

\section{Conclusions} \label{Conclusions}

Observational data supports the idea that galaxies in clusters are
substantially different from galaxies in the field. There is a clear
correlation between galaxy morphology and density of the local
environment. However, the origin of this relation is far from being
understood.

In this study we have presented an analysis of satellite
galaxies that once passed through the virial radius close to the
centre of their respective host halo, but are now found outside the
virial radius in the outskirts of the cluster. We have shown that this
backsplash population is not negligible and needs to be accounted for
when interpreting the various galaxy morphology relationships and
decoupling the degeneracy between nature and nurture.

We must also appreciate that the infalling population is not
expected to be pristine. Rather, we would expect that infall galaxies
have undergone some sort of pre-processing in groups before entering
the cluster too as indicated by the sub-subhalos in \Fig{RminRz}.

Our results can be summarized as follows:

\begin{itemize}

 \item 30\% of all galaxies that ever came closer to the host than its
       virial radius are now located in the range
       [$R_{\rm vir}$,2.5$R_{\rm vir}$],

 \item 50\% of all galaxies in the region [$R_{\rm vir}$,2$R_{\rm vir}$]
       are backsplash galaxies,

 \item 90\% of the backsplash galaxies penetrated deeper than 50\% of 
       $R_{\rm vir}$ into the host's potential,

 \item during their passage through the cluster, on average the backsplash galaxies lose 40\% of their mass, thus

 \item the mass spectrum of the backsplash population has a steeper
       power-law slope than their infalling counterparts, thus it has fewer massive
galaxies and more light ones,

 \item the velocities of the infalling satellites is too small
       to account for ram-pressure stripping in the cluster outskirts,

 \item the backsplash population has a factor of two smaller relative
       velocity than the infalling satellites, making it
       kinematically distinct.

\end{itemize}

When transforming the last result into the observers plane though, the
velocity separation between the infalling and backsplash population is
removed. However, the backsplash population should still be detectable
as it is responsible for a continuous rise in the distribution
function towards low line-of-sight velocities.

Our results suggest that we not only expect the backsplash
population to experience various large-scale transformation
mechanisms, but also small-scale ones, undergoing starvation, ram
pressure stripping, tidally triggered star formation and significant
tidal stripping.

\section{Acknowledgments} 
SPDG wishes to thank Erica Ellingson and Bernard Vollmer for useful discussions. The simulations presented in this paper were carried out on the Beowulf cluster at the Centre for Astrophysics~\& Supercomputing, Swinburne University. The financial support of the Australian Research Council is also gratefully acknowledged. Finally, we wish to thank Michael Balogh for helpful correspondences. 


\end{document}